\documentclass[9pt,twocolumn,twoside]{osajnl}

\journal{jocn} 

\setboolean{shortarticle}{false}
\usepackage{physics}      % \ket, \bra, \bra{}{} etc.
\usepackage{adjustbox}
\usepackage{dblfloatfix}
\usepackage{ragged2e}
\usepackage{makecell}
\usepackage{verbatim} % for comments only, can delete for final version
\newcommand{\hide}[1]{}

\title{Quantum Teleportation of Polarization States on Live Metropolitan Fiber}

\author[1]{Zofia A. Borowska}
\author[2]{Shane Andrewski}
\author[3]{Giorgio De Pascalis}
\author[2]{Olivia Brasher}
\author[2]{Alexander N. Craddock}
\author[2]{Mael Flament}
\author[3]{Klaus D. J\"ons}
\author[2]{Niccol\`o Bigagli}
\author[1]{Ronny D\"oring}
\author[1]{Michaela Ritter}
\author[4]{Ralf-Peter Braun}
\author[1]{Marc Geitz}
\author[1]{Oliver Holschke}
\author[1,*]{Matheus Sena}
\author[2]{Mehdi Namazi}

\affil[1]{Deutsche Telekom AG, Winterfeldtstra\ss e 21, 10781 Berlin, Germany}
\affil[2]{Qunnect Inc, 141 Flushing Ave, Suite 1110, Brooklyn, NY 11205, USA}
\affil[3]{Institute for Photonic Quantum Systems (PhoQS), Center for Optoelectronics and Photonics Paderborn (CeOPP) and Department of Physics, Paderborn University, Warburgerstra\ss e 100, 33098 Paderborn, Germany}
\affil[4]{Orbit GmbH, Mildret-Scheel-Stra\ss e 1, 53175 Bonn, Germany}

\affil[*]{matheus.ribeiro-sena@telekom.de}

\begin{abstract}
As quantum technologies mature, telecommunication operators have a clear opportunity to unlock and scale new services by providing the connectivity layer that links quantum computers, sensors, clocks, and other quantum devices. Realizing this opportunity requires demonstrating quantum networking protocols, including quantum teleportation, under real-world conditions on existing telecom infrastructure. In this work, we demonstrate quantum teleportation deployed on Deutsche Telekom's metropolitan fiber testbed in Berlin using commercial components. A local Bell-state measurement between a weak coherent state at 795~nm and a 795~nm entangled photon enables conditional polarization state transfer onto the 1324~nm (O-band) partner photon, which is transmitted through a 30~km field-deployed fiber. The teleported polarization state of the O-band photon is reconstructed after propagation via quantum state tomography, achieving an average teleportation fidelity of $>$90\% on the deployed link. System performance is evaluated in both the absence and presence of co-propagating C-band classical traffic within the same fiber, demonstrating compatibility with wavelength-division multiplexed telecom infrastructure carrying live classical data channels.
\end{abstract}

\setboolean{displaycopyright}{false} % Do not include copyright or licensing information in submission.

\begin{document}

\maketitle

\section{Introduction}
Quantum teleportation \cite{Bennett95} is a fundamental primitive for quantum networking \cite{kimble2008quantum}, utilizing shared entanglement to transfer an arbitrary qubit state from a sender to a remote receiver without physically transmitting the carrier itself. Teleportation underpins distributed quantum computing \cite{main2025distributed}, remote quantum sensing \cite{jahromi2022remote}, and quantum-repeater architectures \cite{briegel98prl}. In future quantum networks, telecommunications operators are expected to deploy and manage these capabilities much the same way they currently provide classical connectivity, making teleportation a core network mechanism rather than a standalone experiment. Therefore, realizing this primitive on existing operator fiber infrastructure is a prerequisite to offer quantum connectivity as a network service.

In fiber-based quantum networking, quantum information is commonly encoded into photons using degrees of freedom like polarization, time-bin, frequency-bin, or energy-time. Polarization encoding is particularly attractive because it is straightforward to generate, manipulate, and measure, while also interfacing naturally with many matter-based quantum memories and processors. However, its transmission over optical fiber is highly sensitive to time-varying birefringence, resulting from temperature fluctuations, mechanical stress, and vibration, which in turn induce polarization drifts \cite{sena25OFC}.

% Requires \usepackage[table]{xcolor}
\nocite{thomas2024quantum,Rieser2025ECOC_OBandEntanglement,FanPRA23,Craddock24PRX} % to keep citation numbering constant with table
\renewcommand{\arraystretch}{1.05}

\begin{table*}[!b]
\centering

% Very faint gray background for all body rows
\rowcolors{2}{gray!3}{gray!3}

\begin{adjustbox}{width=\textwidth}
\begin{tabular}{|c|c|c|c|c|c|c|c|}
\hline
\rowcolor{gray!10}
\textbf{Reference} &
\textbf{Year} &
\textbf{Setting} &
\textbf{Encoding} &
\textbf{Distance, $\mathrm{km}$} &
\textbf{Fidelity, $\mathrm{\%}$} &
\makecell{\textbf{Classical} \\ \textbf{capacity, $\mathrm{Gbps}$}} &
\makecell{\textbf{Classical} \\ \textbf{power, $\mathrm{dBm}$}} \\
\hline

\cite{gisin2003longdistance} & 2003 & Laboratory & Time-bin & 2 & 81.2 & dark & N/A \\
\cite{bussiere2014telecom} & 2014 & Laboratory & Polarization & 12.4 & 81 & dark & N/A \\
\cite{takesue2015quantum} & 2015 & Laboratory & Time-bin & 102 & 83.7 & dark & N/A \\
\cite{valivarthi2016quantum} & 2016 & Urban & Time-bin & 6.2 & 78 & dark & N/A \\
\cite{sun2016quantum} & 2016 & Urban & Time-bin & 30 & 85 & dark & N/A \\
\cite{valivarthi2020Teleportation} & 2020 & Laboratory & Time-bin & 22 & 89 & dark & N/A \\
\cite{lagorivera2023longdistance} & 2023 & Laboratory & Time-bin & 1 & 86 & dark  & N/A \\
\cite{shen2023hertz} & 2023 & Laboratory & Time-bin & 64 & 90.6 & dark & N/A \\
\cite{kucera2024demonstration} & 2024 & Urban & Polarization & 14.4 & 78--86 & dark & N/A \\
\cite{thomas2024quantum} & 2024 & Laboratory & Polarization & 30.2 & 89.9 & 400 & 18.7 \\
\cite{song2026hollowcore} & 2026 & Urban (HCF) & Time-bin & 15.3 & 86.2 & - & 22.0 \\
\rowcolor{pink!30}
This work & 2026 & Urban & Polarization & 30 & 90.4 & 100 & 0.27 \\

\hline
\end{tabular}
\end{adjustbox}
\caption{Comparison of performance metrics of photonic fiber-based quantum teleportation on comparable testbeds, specifically fiber type, distance, fidelity, encoding of the teleported channel (\textit{polarization/time-bin}), and if available, co-propagation details like classical capacity and power. We distinguish two types of experiments: fiber spool-based in-lab setups or nearby buildings supplemented with spools (\textit{laboratory}), and real-world deployments using installed fiber (\textit{urban}). HCF denotes hollow-core fiber; all other experiments use standard single-mode fiber (SMF). Experiments without co-propagating classical traffic are identified as operating under dark-fiber conditions, with \textit{dark} listed for classical capacity and \textit{N/A} for the corresponding power. Distances are quoted as reported and follow differing conventions (see footnote).}
\label{tab:RelatedWork}
\end{table*}

Compared to classical communication, quantum state transmission thus imposes stricter requirements on channel stability and noise, but existing telecommunication infrastructure has not been designed to preserve quantum polarization states. Moreover, many practical applications require quantum and classical signals to coexist within the same fiber \cite{thomas2024quantum} to enable efficient reuse of commercial components and deployed infrastructure. When quantum and classical signals co-propagate, inter-channel crosstalk, dominated by Raman scattering, introduces excess noise that degrades quantum correlations irrespective of encoding \cite{Rieser2025ECOC_OBandEntanglement, FanPRA23}. Addressing these challenges, i.e., polarization drifts and inter-channel crosstalk, requires both automatic polarization compensation to correct environmentally induced polarization rotations \cite{Craddock24PRX} and wavelength-division multiplexing (WDM) to suppress classical noise.

Fiber-based quantum teleportation demonstrations, summarized in \autoref{tab:RelatedWork} \cite{song2026hollowcore, liu2025chip, thomas2024quantum, kucera2024demonstration, shen2023hertz, lagorivera2023longdistance, valivarthi2020Teleportation, sun2016quantum, valivarthi2016quantum, takesue2015quantum, bussiere2014telecom, gisin2003longdistance}, have matured from early proof-of-concept local experiments to longer-distance demonstrations\footnote{A note on distance conventions: the most common convention, and the one we adopt for this work, is the state-transfer distance where the propagation length of the teleported photon is from the Bell-state measurement to the receiver. A smaller number of experiments instead report the combined path of the two photons entering the BSM, or the straight-line node separation of the communicating nodes. We quote each value as reported in its original work, which may follow any of these conventions.}. However, the highest-performing results are typically obtained on laboratory fiber spools under controlled conditions, where polarization fluctuations and environmental drift are minimal. Relatively few of these experiments are conducted in installed metropolitan fiber, and fewer still report co-propagation with live classical traffic (see \autoref{tab:RelatedWork}). Our work addresses this gap, demonstrating polarization-based teleportation on installed metropolitan fiber, while sharing the link with live telecom traffic.

%Additionally, for teleportation to serve as a practical network primitive, it must interface with the quantum devices that form the network's end nodes. To address this, we work in two regimes. First, we realize bichromatic teleportation, using a wavelength-asymmetric interface that links 795-nm light, compatible with rubidium-based devices, to the telecom O-band for low-loss transmission. Second, we prepare the teleported input with a commercially available attenuated laser, representative of the off-the-shelf hardware a real network node would use. 

% For teleportation to serve as a practical network primitive, it must also interface with the quantum devices that form the network's end nodes. Many such devices, including trapped ions, neutral atoms, atomic memories, clocks, and sensors, operate at optical transitions far from the telecom band where long-distance transmission is most efficient. Bridging these regimes calls for wavelength-asymmetric interfaces, in which the locally processed photon is matched to a matter qubit while the transmitted photon is matched to the fiber. We realize such an interface with a bichromatic polarization-entangled source that links 795-nm light, compatible with rubidium-based devices, to the telecom O-band. The teleported input state is likewise prepared with a commercially available attenuated laser, emulating the variety of input sources a practical quantum network node could employ. Both choices reflect the off-the-shelf hardware from which a practical quantum network would realistically be built.

In a previous work, we demonstrated high-fidelity entanglement distribution over Deutsche Telekom's deployed metropolitan fiber network \cite{sena25JOCN}. Here, we extend that platform to quantum teleportation. A time-resolved Bell-state measurement (BSM) is performed locally between a weak coherent state (WCS) and the 795~nm photon of a non-degenerate polarization-entangled pair. Conditional on a successful BSM, the input polarization state of the WCS is teleported, up to a known unitary rotation, onto the 1324~nm O-band photon which is transmitted through a 30-km deployed metropolitan fiber link. Employing automatic polarization compensation and WDM, measurements are performed in both a dark-fiber configuration and in the presence of a co-propagating 100~Gbps dense wavelength-division multiplexing (DWDM) signal in the C-band, the latter realizing coexistence on a live link.

\begin{figure*}[!t]
 \centering
 \includegraphics[width=1\textwidth]{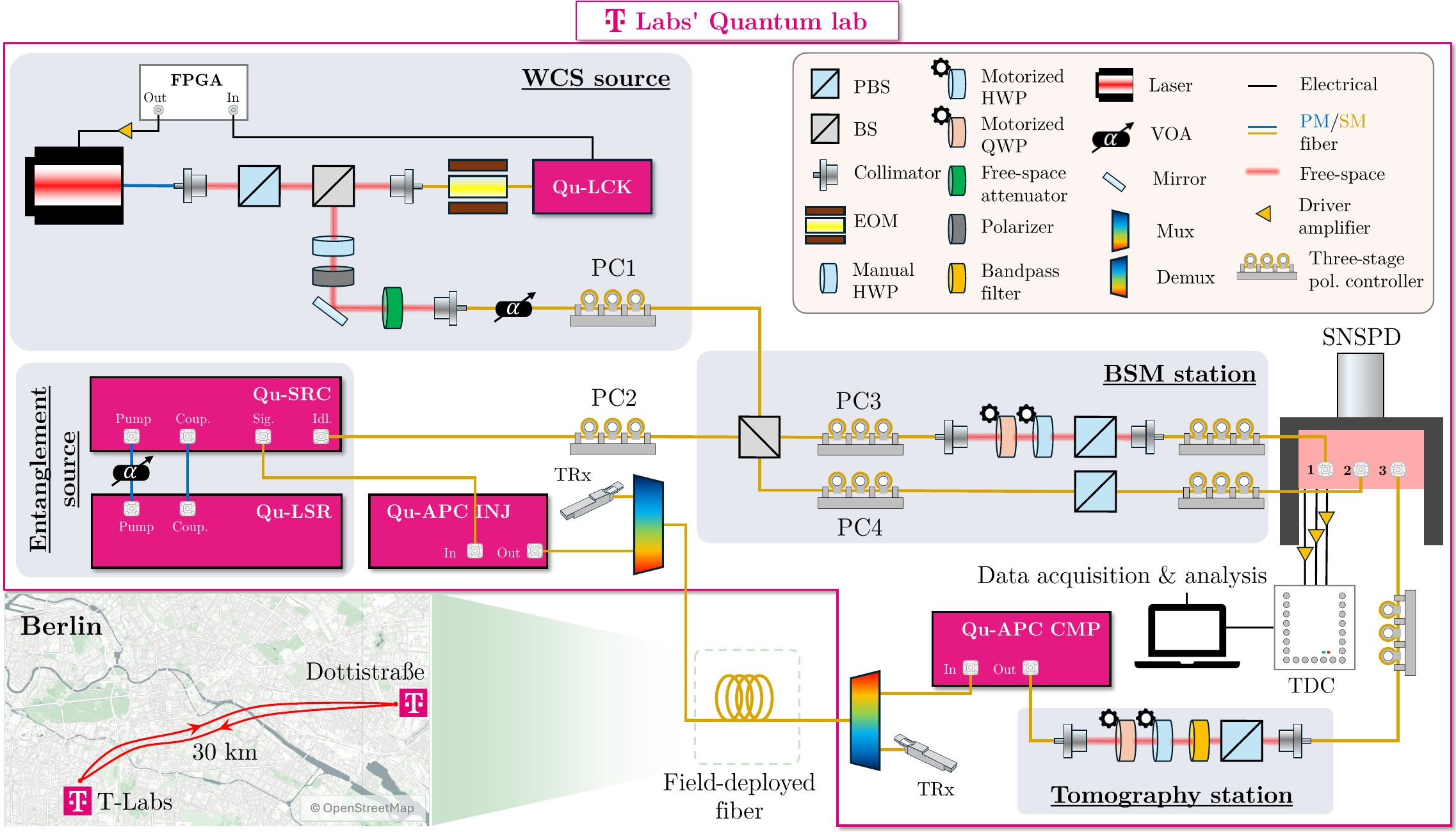}
 \caption{Schematic of the field-deployed teleportation experiment on Deutsche Telekom's R\&D fiber test network in Berlin. A 795~nm weak coherent state (WCS) interferes with the 795~nm photon of a bichromatic polarization-entangled pair at a local Bell-state measurement (BSM). Conditional on a successful BSM, the input polarization state is transferred, up to a known unitary, onto the 1324~nm photon, which is transmitted over a 30~km deployed fiber and analyzed by polarization tomography. Co-propagating C-band classical traffic is multiplexed onto the same fiber as the 1324 nm photon.
 }
 \label{fig:Hardware_Diagram}
\end{figure*} 

\section{Theoretical Description}
\label{sec:theory}
% ---------------------------------------------------------
In this section, we relate the teleportation fidelity to two independent quantities, the entanglement fidelity of the pair source and the photon indistinguishability at the BSM.
% so that the protocol's performance can be verified directly against experimental data.

% Our teleportation protocol combines three components: an entangled photon pair from a spontaneous four-wave-mixing (SFWM) source, an input qubit  (WCS), and a BSM that jointly projects the input photon and one photon of the entangled pair onto a Bell state. A successful BSM projects the state of the input photon onto the remaining heralding photon of the SFWM source, up to a known unitary operation. 
% The quality of the protocol is limited by two imperfections: the degradation of the entangled source state, and an imperfect BSM. We describe each of these separately, characterizing the source by its entanglement fidelity and the BSM by the photon indistinguishability, and combine them to obtain the teleportation fidelity.

First, we express the environmental influence on the $\ket{\Phi^+}$ entangled state through the density matrix
\begin{equation}
    \label{eq:sfwm_state}
    \rho_{\mathrm{pair}} = (1 - p)\ket{\Phi^+}\bra{\Phi^+} + \frac{p}{4}\,\mathbb{1},
\end{equation}
which mixes the ideal Bell state with an unpolarized background (represented by identity $\mathbb{1}$), weighted by $p$. This single parameter includes the effects of multi-photon emission from the source as well as its interactions with the experimental setup. In this description, the entanglement fidelity of the source is then:
\begin{equation}
    \label{eq:sfwm_ent_fidelity}
    F_e = \bra{\Phi^+}\rho_{\text{pair}}\ket{\Phi^+} = 1 - \frac{3}{4}p.
\end{equation}

Second, we consider imperfections of the BSM.  In our scheme, only the two cross-polarized states $\ket{\Psi^\pm}$ can be resolved. 
% These two outcomes are told apart by Hong-Ou-Mandel (HOM) interference \cite{hong1987hom}, which cleanly separates them only when the interfering WCS and SFWM photons are indistinguishable. 
We quantify the similarity of the photons by a parameter, $0\leq c\leq 1$ \cite{craddock2019}, where $c=1$ represents perfect indistinguishability and $c=0$ perfect distinguishability. The probability that the state is correctly heralded follows from this parameter,
\begin{equation}
    \label{eq:bsm_posteriors}
    P(\Psi^-\,|\,\Psi^-_\mathrm{BSM}) = \frac{1+c}{2}, \qquad
    P(\Psi^+\,|\,\Psi^-_\mathrm{BSM}) = \frac{1-c}{2}.
\end{equation}

Combining the two effects, the heralded output is a mixture of these two arrivals: with weight $(1+c)/2$ the target state is reproduced (degraded only by the source noise $p$), and with weight, $(1-c)/2$, an incorrect state is produced. Evaluating the fidelity and averaging over the mutually unbiased inputs $\{\ket{H},\ket{D},\ket{R}\}$ gives the relation,
\begin{equation}
    \label{eq:avg_teleportation_fidelity}
    \overline{F}_T
        = \frac{2}{3} - \frac{p}{6} + \frac{1-p}{3}\,c,
\end{equation}
which, substituting $p = \tfrac{4}{3}(1-F_e)$, becomes:
\begin{equation}
    \label{eq:avg_teleportation_fidelity_measured}
    \overline{F}_T
        = \frac{2+c}{3} - \frac{2}{9}(1-F_e)(1+2c),
        \qquad c,F_e \in [0,1],
\end{equation}
% visualized in \autoref{fig:teleportation_fidelity_theory}.

% \begin{figure}[ht]
%     \centering
%     \includegraphics[width=1\columnwidth]{figures/teleportation_fidelity_theory.pdf}
%     \caption{Average teleportation fidelity $\overline{\mathcal{F}}_T$ as a function of the entanglement fidelity $\mathcal{F}_e$ and the photon similarity, $c$. The markers show our three measured operating points: local (circle), 30~km dark (triangle), and 30~km live (square). The dashed line is the classical bound $\overline{\mathcal{F}}_T = 2/3$; everything below is shown in black.}
%     \label{fig:teleportation_fidelity_theory}
% \end{figure}

A perfect source ($F_e = 1$) and perfect interference ($c = 1$) give $\overline{F}_T = 1$, exact teleportation, while a fully mixed source ($F_e = 1/4$) gives $\overline{F}_T = 1/2$, carrying no information about the input. For $c=0$, $\overline{F}_T=2/3$. 

\section{Experimental Setup and Methods}
\label{sec:exp_setup}

In this section, we detail the experimental setup illustrated in \autoref{fig:Hardware_Diagram}, together with  measurement procedures. 

\subsection{Weak Coherent State}
The input polarization states to be teleported are prepared by an attenuated continuous-wave (CW), frequency-locked, 795~nm Distributed Bragg Reflector (DBR) laser. For locking, the laser light is phase-modulated with an electro-optic modulator (EOM) and sent to a rubidium (Rb) spectroscopy module (Qu-LCK). The absorption signal is processed by an FPGA to derive an error signal. The frequency-locking reference is chosen such that the laser's wavelength is overlapped with the  795~nm entangled photon frequency spectrum. 

The light is coarsely attenuated with neutral-density filters and fine-tuned with a fiber-based voltage-controlled optical attenuator (VOA) to reach the  single-photon level while retaining control of the single-photon rate. A polarizing beam splitter (PBS) provides the initial polarization stabilization, and a motorized three-stage polarization controller (PC1) then prepares the input state for teleportation. The detected single photon rate of the WCS is $13\times10^6/s$ summed across channels 1 and 2 (in a 32~ps coincidence window). 

%The detected rates from the WCS are 5.8~Mhz for channel~1 and 7.6~MHz for channel~2, and the transmission efficiency from BS to the detector are 2.6\% (channel~1) and 4.0\% (channel~2).

% Using an attenuated laser in place of a dedicated single-photon source provides a simple, tunable, and commercially available transmitter representative of the input sources a practical network node could employ, while its residual multi-photon component constitutes a realistic impairment \cite{xu2023characterization}. %that is directly reflected in the measured HOM visibility and the resulting teleportation fidelities

\subsection{Entanglement Source}
Polarization-entangled photon pairs are generated with Qunnect's Qu-SRC. Non-degenerate photon pairs at 795~nm and 1324~nm are emitted from a warm $^{87}\mathrm{Rb}$ ensemble via spontaneous four-wave mixing (SFWM) in a diamond configuration \cite{PhysRevApplied.21.034012}. Pump (780~nm) and coupling (1367~nm) fields, provided by the Qu-LSR laser system, drive the SFWM process.  

The emitted state approximates $\ket{\Phi^{+}}=(\ket{HH}+\ket{VV})/\sqrt{2}$. At the source output, 795~nm photons pass through a motorized polarization controller (PC2) toward the BSM station, while 1324~nm photons are routed either locally or through the metropolitan fiber. The detected 795~nm photon rate is $530\times10^3/s$ singles counts across channels 1 and 2. The detected 1324~nm photon rate is $640\times10^3/s$ in local configuration, and the $g_{si}$ %measured from coincidences $1\otimes3$ 
is 315. At this rate we measure the SFWM entanglement fidelity as $F_e = 0.966$ using a full two-photon tomography measurement.

%The detected 795~nm rates are 190~kHz on channel 1 and 340~kHz on channel 2. The detected 1324~nm rate on channel is 640~kHz (in local configuration). Within a 32~ps coincidence window, the channel~1$\otimes$3 coincidence rate is 1.2~kHz, corresponding to $g_{si}\approx315$.

% The source can achieve pair rates exceeding \(10^{7}~\mathrm{s^{-1}}\), Bell-state fidelity \(>95\%\), sub-GHz biphoton linewidth (below \(2\pi\times1~\mathrm{GHz}\)), spectral brightness of \(\sim10^{4}~\mathrm{s^{-1}\,MHz^{-1}}\), and strong correlations with \(g_{si}\ge50\) \cite{PhysRevApplied.21.034012}.

\subsection{Bell State Measurement Station}
The WCS photon and the $795~\mathrm{nm}$ photon from the Qu-SRC interfere at a fiber-based 50:50 non-polarizing beam-splitter (NPBS) in a local BSM. The two outputs are connected to motorized polarization controllers (PC3 in the upper path and PC4 in the lower path; see \autoref{fig:Hardware_Diagram}) used for calibration. In the upper path, the photons pass through a free-space motorized QWP and HWP and a free-space PBS, while in the lower path, they pass through a fiber-based PBS only. The two PBS outputs are routed to channels 1 and 2 of the Single Quantum Superconducting Nanowire Single-Photon Detector (SNSPD) system (detection efficiencies 82\%, 90\%, and 90\% for channels 1, 2, and 3; jitter $\leq$18~ps), with a manual polarization controller on each channel aligning the light to the detector's preferred axis. Detection events are time-tagged by a Swabian Instruments time-to-digital converter (TDC) and processed on a data acquisition computer.

\subsection{Field-Deployed Fiber Loop and Stabilization}
\begin{figure}[h!]
 \centering
 \includegraphics[width=1\columnwidth]{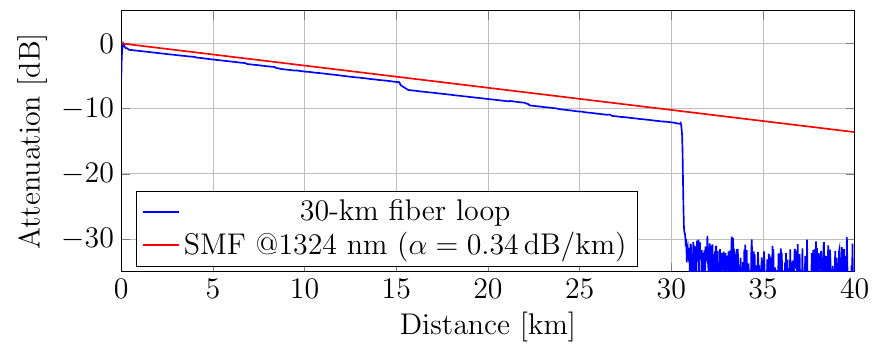}
 \caption{OTDR trace of the 30-km deployed fiber used in this experiment (T-Labs to Dottistra\ss e). Discontinuities correspond to connectors, splices, or patch-panel events. The red line indicates the expected attenuation of standard SMF-28 at 1324~nm (approximately 0.34~dB/km).}
 \label{fig:otdr}
\end{figure}

We characterize the field-deployed fiber with an optical time-domain reflectometer (OTDR), as installed fibers exhibit splice and coupler losses, and additional attenuation from manufacturing or installation inconsistencies. The selected loop between T-Labs' Quantum Lab and Deutsche Telekom's office on Dottistra\ss e shows losses close to expected values, with an attenuation profile (blue curve, \autoref{fig:otdr}) that closely matches standard single-mode fiber (SMF) at 1324~nm (red curve).

To maintain stable operation over deployed fiber, the O-band 1324~nm photon is routed through paired Qu-APC Injector and Compensator modules, placed before and after the loop, which use time-multiplexed classical test signals to actively compensate fiber-induced polarization rotations \cite{Craddock24PRX, Flament2025AU2021365730B2}. Triggered at predefined intervals ($>$1~h) during long acquisitions, each $\sim$1~s compensation achieves classical polarization fidelities of 99.5--99.9\%, ensuring the 30~km fiber channel does not degrade the teleported state despite gradual drift \cite{sena25OFC}. The modules also enable a calibration strategy that circumvents loss limitations: we calibrate locally on a short low-loss fiber and then switch to the 30~km link, letting the modules compensate for the new fiber transformation \cite{Rieser2025ECOC_OBandEntanglement}. Between the Injector and the Compensator, a multiplexer (Mux) combines the O-band photon with the C-band classical channel for transmission through the loop, and a demultiplexer (Demux) separates them on return before the Compensator and tomography station. The total end-to-end attenuation from the Qu-APC INJ output to the Qu-APC CMP input is -12~dB, including deployed-fiber loss, Mux/Demux insertion loss, and connector and patch-panel losses.

\subsection{Co-propagating Classical Channel}
To assess coexistence with live telecom traffic, a pair of optical transceivers (TRx) are connected directly to the Mux/Demux system. The classical channel is provided by a 100G DWDM transmitter centered at 1561.01~nm at a fixed output power of 0.27~dBm. A 12~nm bandpass filter inserted between the HWP and the PBS at the tomography station partially suppresses spurious light from this channel in channel 3.

\subsection{Tomography Station}
The tomography station reconstructs the teleported polarization state through projective measurements, from which its fidelity to the target state is evaluated (as described in the following subsection). The station provides fiber-coupled input and output with free-space motorized polarization analysis implemented with a QWP, HWP, and PBS. The 1324~nm photons are detected on channel 3 of the SNSPD system (with a manual polarization controller aligning the incoming light to the detector's preferred axis), and detection events time-tagged by the TDC.

\subsection{Teleportation Protocol and State Reconstruction}
\label{ssec:protocol}

Teleportation events are identified through three-fold coincidence measurements conditioned on successful two-photon $\ket{\Psi^-}$ BSM outcomes, heralded by detection of the 1324~nm photon at the tomography station, within the 32~ps coincidence window.
% Because only one output port of each PBS is monitored, the experiment is sensitive exclusively to $\ket{\Psi^-}$ projections.

The teleported state is reconstructed by polarization tomography in the rectilinear, diagonal, and circular bases. The QWP and HWP are set to six configurations corresponding to projections onto $\{\ket{H}, \ket{V}, \ket{D}, \ket{A}, \ket{R}, \ket{L}\}$.
% , with the angles given in \autoref{tab:state_tomography}. 

% \begin{table}[htbp]
% \centering
% \caption{\bf QWP and HWP relative rotation angles used in the tomography station to project onto the polarization basis states $\{\ket{H}, \ket{V}, \ket{D}, \ket{A}, \ket{R}, \ket{L}\}$.}
% \setlength{\tabcolsep}{6pt}
% \renewcommand{\arraystretch}{1.1}
% \begin{tabular}{c|cc}
% \hline
% State & QWP (\textdegree) & HWP (\textdegree) \\
% \hline
% $\ket{H}$ & 0 & 0 \\
% $\ket{V}$ & 0 & 45 \\
% $\ket{D}$ & 45 & 22.5 \\
% $\ket{A}$ & 45 & 67.5 \\
% $\ket{R}$ & 45 & 0 \\
% $\ket{L}$ & 0 & 22.5 \\
% \hline
% \end{tabular}
% \label{tab:state_tomography}
% \end{table}

The Stokes parameters are obtained from the measured threefold coincidence counts $N$ in the corresponding polarization bases as:
\begin{equation}
\begin{pmatrix}
S_x & S_y & S_z
\end{pmatrix}
=
\begin{pmatrix}
\dfrac{N_D - N_A}{N_D + N_A} &
\dfrac{N_R - N_L}{N_R + N_L} &
\dfrac{N_H - N_V}{N_H + N_V}
\end{pmatrix}.
\end{equation}
% From these, we find the state density matrix $\rho$ (\autoref{eq:rho_from_stokes}), and determine the fidelity $F$ to the target states $\ket{\psi_T} = \{\ket{H}, \ket{D}, \ket{R}\}$ (\autoref{eg:fidelity_from_rho}).

\noindent From these we find the density matrix $\rho$ of the teleported state
\begin{equation}
\rho
= \frac{1}{2}\Bigl(
I
+ S_x\,\sigma_x
+ S_y\,\sigma_y
+ S_z\,\sigma_z
\Bigr),
\end{equation}
where $I$ denotes the identity operator and $\sigma_i$ $(i = x,y,z)$ are the Pauli matrices. Finally, the fidelity $F$ of the teleported state with respect to the target state is then evaluated as follows:
\begin{equation}
F = \bra{\psi_T}\rho\ket{\psi_T},
\end{equation}
where $\ket{\psi_T}$ denotes the target polarization state. The fidelity uncertainty is estimated by Monte Carlo simulation: threefold coincidence counts are sampled from a Poissonian distribution \cite{james2001measurement} and the fidelity calculation is repeated over 10,000 trials to obtain the standard deviation of the resulting distribution. 
% \autoref{tab:state_mapping} provides the correspondence between the teleported state and the consequently received state given the BSM detection conditioned on $\ket{\Psi^{-}}$, and no active-feed forward for the final unitary transformation, for the three input states used to demonstrate quantum teleportation.

% \begin{table}[htbp]
% \centering
% \caption{\bf State mapping conditioned on the $\ket{\Psi^-}$ Bell-state measurement outcome. No corrective unitary is applied.}
% \setlength{\tabcolsep}{6pt}
% \renewcommand{\arraystretch}{1.1}
% \begin{tabular}{c|c}
% \hline
% Teleported & Received ($\ket{\Psi^-}$) \\
% \hline
% $\ket{H}$ & $\ket{V}$ \\
% $\ket{D}$ & $\ket{A}$ \\
% $\ket{R}$ & $\ket{R}$ \\
% \hline
% \end{tabular}
% \label{tab:state_mapping}
% \end{table}

The three-fold coincidence rate was of order $1$~Hz in the local configuration, dropping to $\sim$0.1~Hz over the 30~km link. For sufficient statistics in each basis, data was acquired for 180~s per setting locally, 1200~s per setting over the 30~km dark fiber, and 1500~s per setting with co-propagating classical traffic (the longer time compensating for the added multiplexing losses). With six projection settings, a full tomographic reconstruction required roughly six times the integration of a single HOM scan at the same rate.

\section{Results and Discussion}
\label{sec:results}

\subsection{Hong--Ou--Mandel Interference}

\begin{figure}[htb]
 \centering
 \includegraphics[width=1\columnwidth]{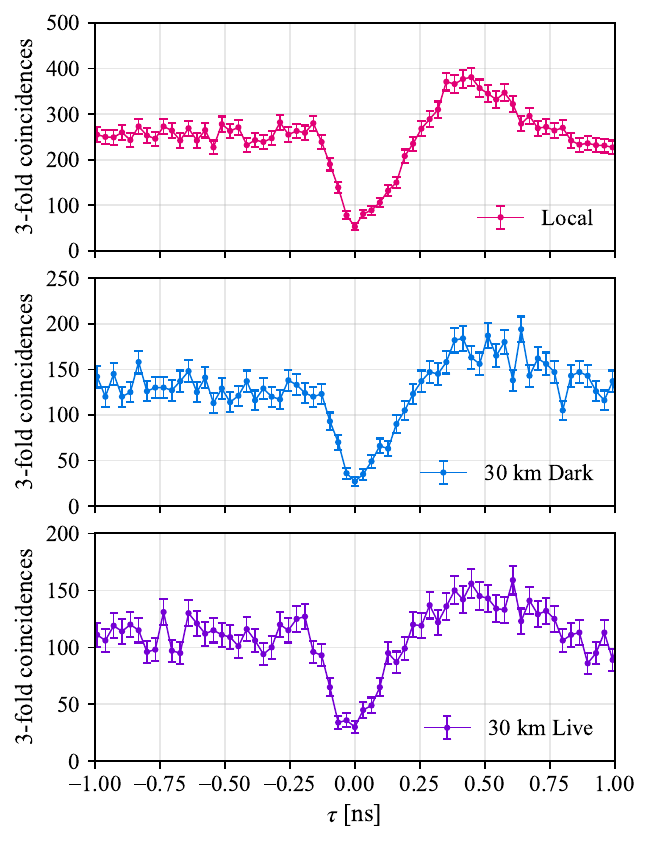}
 \caption{Hong--Ou--Mandel interference for the local (magenta, top), 30~km dark (blue, middle), and 30~km live (purple, bottom) configurations. Three-fold coincidences (unnormalized) between the two NPBS outputs and the 1324~nm heralding photon are plotted against the relative delay $\tau$. Coincidence window is 32~ps. Error bars denote Poissonian uncertainties.}
 \label{fig:hom}
\end{figure}

We characterize the time resolved Hong--Ou--Mandel (HOM) interference between the WCS and the Qu-SRC 795~nm photons by measuring three-fold coincidences as a function of the relative temporal delay $\tau$ between the two input photons. The three-fold coincidences are formed from the two NPBS outputs and the heralding 1324~nm photon, and the visibility is calculated as $V=(N_{\tau=\infty}-N_{\tau=0})/N_{\tau=\infty}$, where $N_{\tau=0}$ denotes the coincidence counts at zero delay ($\tau=0$), and $N_{\tau=\infty}$ is obtained from the mean coincidence counts far from the dip ($\tau=-10$ to $-5$~ns), where the photons are fully temporally distinguishable. 
We note that in this experimental configuration the visibility may take values $-1\leq V \leq 1$, and is related to the photon similarity by $V=2c-1$.
Results for the three configurations are shown in \autoref{fig:hom}.

The measured visibilities are $V_{\mathrm{local}} = 78.9 \pm 3.2\%$ for the local configuration and $V_{\mathrm{30\,km,\,dark}} = 78.6 \pm 4.6\%$ after transmission through the 30-km dark fiber, leaving the interference essentially unchanged by long-distance transmission. With co-propagating classical traffic, the visibility drops to $V_{\mathrm{30\,km,\,live}} = 72.8 \pm 5.7\%$, reflecting the additional background introduced on the heralding channel.

\subsection{Teleportation Results}

\begin{figure}[ht]
 \centering
 \includegraphics[width=1\columnwidth]{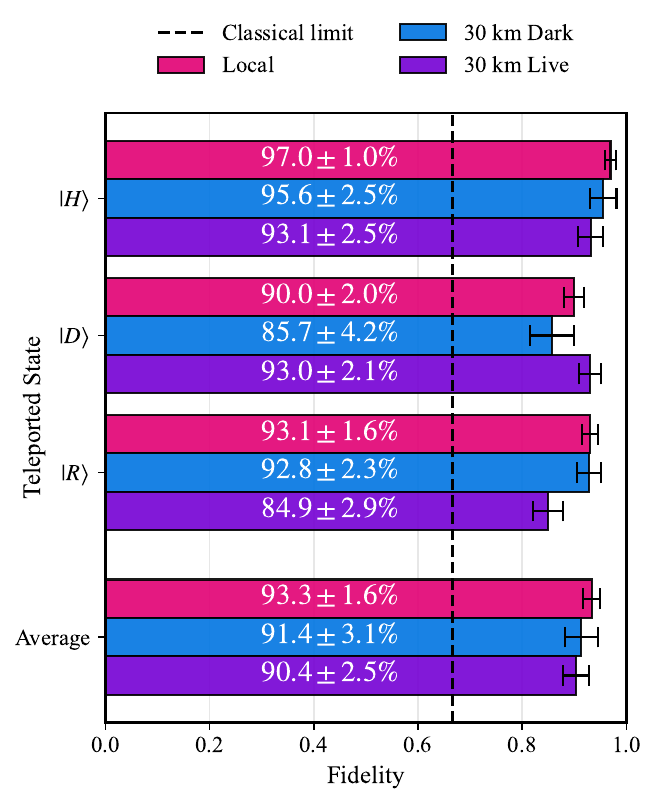}
 \caption{Teleportation fidelities for the teleported states $\{ \ket{H},\ket{D},\ket{R}\}$: local (magenta, left), 30~km deployed fiber
without co-propagation (blue, center), and 30 km with co-propagating classical traffic (purple, right). Rightmost bars show the average fidelity for each configuration. The dashed line indicates the classical bound $F=2/3$.}
 \label{fig:fidelity_barchart}
\end{figure}

\begin{figure*}[!ht]
 \centering
 \includegraphics[width=1\textwidth]{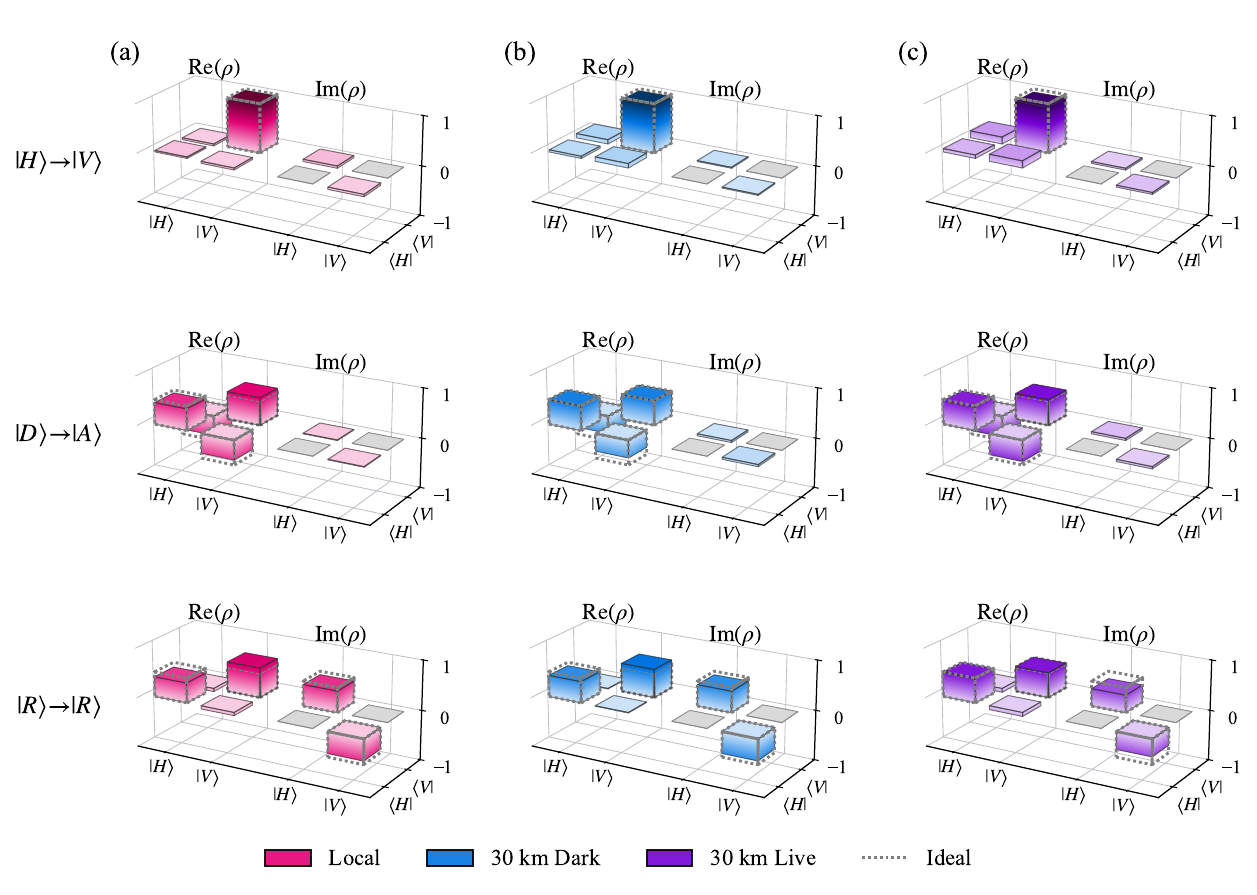}
 \caption{Reconstructed density matrices $\rho$ for the teleported output states. Each column corresponds to one of the three configurations: (a) local (magenta), (b) 30~km dark (blue), and (c) 30~km live (purple), and the rows represent the teleported states. The measured output states ${\ket{V},\ket{A},\ket{R}}$ correspond to the input states ${\ket{H},\ket{D},\ket{R}}$, as expected in the absence of the unitary correction operation in the case of $\ket{\Psi^-}$ conditioning. Each $\rho$ is given in terms of its real (left) and imaginary (right) components in the ${\ket{H},\ket{V}}$ basis. Dotted lines represent the ideal density matrix for each state.}
 \label{fig:tomography_all}
\end{figure*}

\autoref{fig:fidelity_barchart} presents the teleportation fidelities for the input states $\ket{H}$, $\ket{D}$, and $\ket{R}$ under three operating configurations: local, 30~km transmission without co-propagating classical traffic (30~km dark), and 30~km transmission with co-propagating classical traffic (30~km live). Averaged over the three tested states, we obtain fidelities of $93.3 \pm 1.6\%$ (local), $91.4 \pm 3.1\%$ (30-km dark), and $90.4 \pm 2.5\%$ (30-km with co-propagation). All values exceed the classical bound of $F=2/3$ by several standard deviations, confirming teleportation. The reconstructed density matrices for each state and configuration are given in \autoref{fig:tomography_all}.

\subsection{Discussion}

The addition of co-propagating classical traffic introduces additional background counts on detector channel~3, which contribute to the non-negligible off-diagonal elements present in the density matrix in the 30~km live configuration, shown in \autoref{fig:tomography_all}(c). We attribute this excess noise ($\sim30\times10^3/s$) primarily to Raman scattering from the classical channel in the deployed fiber, which lowers the measured entangled-pair cross-correlation from $g_{si} = 315$ to $g_{si} = 192$.

The measured teleportation fidelities are consistent with the theoretical expectation of \autoref{eq:avg_teleportation_fidelity_measured}, which relates the protocol performance to the independently measured entanglement fidelity and HOM visibility. 

The HOM visibility, characterizing the BSM, is constrained by two factors: the residual multi-photon contributions from both the entanglement source and the weak coherent state, and finite detector resolution of the time resolved quantum interference effect \cite{legero2003time}. % BEATING EFFECT HERE?

% Because the link itself is not the limiting factor, the protocol could in principle be extended well beyond 30~km. However, the main cost of greater distance is rate, as when loss accumulates, the integration time required for a given statistical precision grows rapidly. Extending to 60~km, we still observed successful non-classical interference with $V_{\mathrm{60\,km}} = 64.0\pm4.0\%$, but this measurement required two hours integration time, implying order of 12~hours for a full teleportation measurement. Such acquisition times render direct teleportation impractical at these distances with the studied setup, underscoring the need for the rate improvements.

Several routes could improve the rate. The strong signal-idler correlations indicate substantial room to increase the entangled-pair flux through higher pump power. In the longer term, integrating quantum memories at the network nodes would decouple the protocol's success probability from the coincidence of independent photons, through techniques such as multimode storage and on-demand retrieval \cite{haenni25heralded}, providing the most direct route to practical teleportation rates over deployed fiber.

Reducing transmission loss through alternative fiber platforms, such as hollow-core fiber (HCF), could further increase the achievable rate while also supporting co-propagation with high-power classical traffic \cite{song2026hollowcore}. However, widespread adoption of HCF would require new fiber deployment, whereas the goal of this work is to demonstrate quantum teleportation over the existing installed single-mode fiber infrastructure. Although HCF may ultimately become attractive for dedicated quantum infrastructure, demonstrating compatibility with the already-installed SMF network is the more immediate requirement for telecom deployment.

% [COULD BE GOOD TO DISCUSS THE BEATING HERE]

\section{Conclusion}
\label{sec:Conclusion}

We demonstrated quantum teleportation over Deutsche Telekom's metropolitan fiber testbed in Berlin using commercially available, field-deployable quantum hardware and optical components. A Bell-state measurement between a 795 nm weak coherent state and the 795 nm photon of a bichromatic polarization-entangled pair enabled conditional state transfer onto a telecom O-band photon transmitted through a 30-km deployed fiber loop.

Using polarization-state tomography, we reconstructed the teleported states after transmission through the metropolitan network and measured average teleportation fidelities of $91.4 \pm 3.1\%$ without co-propagating classical traffic and $90.4 \pm 2.5\%$ in the presence of a live 100G DWDM C-band channel, both well above the classical limit. Although the classical channel introduced additional Raman background noise, the teleportation fidelity remained largely unchanged, confirming that polarization-based teleportation can coexist with wavelength-division-multiplexed telecom traffic. 

Together with recent demonstrations of entanglement distribution \cite{sena25JOCN} and entanglement swapping \cite{swaperooboubou} over deployed metropolitan networks, this work extends the range of quantum networking primitives that have been realized on existing carrier infrastructure. These results demonstrate that quantum teleportation can operate as a practical network function on existing carrier-grade fiber infrastructure, providing a concrete step toward deployable quantum networking services.

\section*{Funding} 
The authors acknowledge partial financial support from the German Federal Ministry of Research, Technology and Space (BMFTR) under the project $QR.N$ (FKZ 16KIS2181 and 16KIS2206).

% \section*{Acknowledgments}
% \section{References}

% Bibliography
\bibliography{refs}

@article{PhysRevApplied.21.034012,
  title = {High-rate subgigahertz-linewidth bichromatic entanglement source for quantum networking},
  author = {Craddock, Alexander N. and Wang, Yang and Giraldo, Felipe and Sekelsky, Rourke and Flament, Mael and Namazi, Mehdi},
  journal = {Phys. Rev. Appl.},
  volume = {21},
  issue = {3},
  pages = {034012},
  numpages = {6},
  year = {2024},
  month = {Mar},
  publisher = {American Physical Society},
  doi = {10.1103/PhysRevApplied.21.034012},
  url = {https://link.aps.org/doi/10.1103/PhysRevApplied.21.034012}
}

@article{thomas2024quantum,
  title={Quantum teleportation coexisting with classical communications in optical fiber},
  author={Thomas, Jordan M. and Yeh, Fei I. and Chen, Jim Hao and Mambretti, Joe J. and Kohlert, Scott J. and Kanter, Gregory S. and Kumar, Prem},
  journal={Optica},
  volume={11},
  number={12},
  pages={1700-1707},
  year={2024},
  doi={10.1364/OPTICA.540362},
  url={https://opg.optica.org/optica/fulltext.cfm?uri=optica-11-12-1700&id=565936},
}

@article{liu2025chip,
  title={Chip-to-chip photonic quantum teleportation over optical fibers of 12.3 km},
  author={Liu, Dongning and Jin, Zhanping and Liu, Jingyuan and Zou, Xiaotong and Ren, Xiaosong and Li, Hao and You, Lixing and Feng, Xue and Liu, Fang and Cui, Kaiyu and Huang, Yidong and Zhang, Wei},
  journal={Light: Science \& Applications},
  volume={14},
  pages={243},
  year={2025},
  doi={10.1038/s41377-025-01920-z},
  url={https://www.nature.com/articles/s41377-025-01920-z},
}

@article{takesue2015quantum,
  title={Quantum teleportation over 100 km of fiber using highly efficient superconducting nanowire single-photon detectors},
  author={Takesue, Hiroki and Dyer, Shellee D. and Stevens, Martin J. and Verma, Varun and Mirin, Richard P. and Nam, Sae Woo},
  journal={Optica},
  volume={2},
  number={10},
  pages={832-835},
  year={2015},
  doi={10.1364/OPTICA.2.000832},
  url={https://opg.optica.org/optica/fulltext.cfm?uri=optica-2-10-832},
}

@article{sun2016quantum,
  title={Quantum teleportation with independent sources and prior entanglement distribution over a network},
  author={Sun, Qi-Chao and Mao, Ya-Li and Chen, Si-Jing and Zhang, Wei and Jiang, Yang-Fan and Zhang, Yan-Bao and Zhang, Wei-Jun and Miki, Shigehito and Yamashita, Taro and Terai, Hirotaka and Jiang, Xiao and Chen, Xian-Feng and You, Li-Xing and Chen, Xue and Wang, Zhen and Fan, Jing-Yun and Zhang, Qiang and Pan, Jian-Wei},
  journal={Nature Photonics},
  volume={10},
  pages={671--675},
  year={2016},
  doi={10.1038/nphoton.2016.179},
  url={https://www.nature.com/articles/nphoton.2016.179} 
}

@article{valivarthi2016quantum,
  title={Quantum teleportation across a metropolitan fibre network},
  author={Valivarthi, Raju and Grimau Puigibert, Marcel.li and Zhou, Qiang and Aguilar, Gabriel H. and Verma, Varun B. and Marsili, Francesco and Shaw, Matthew D. and Nam, Sae Woo and Oblak, Daniel and Tittel, Wolfgang},
  year={2016},
  journal={Nature Photonics},
  volume={10},
  pages={676--680},
  doi={https://doi.org/10.1038/nphoton.2016.180},
  url={https://arxiv.org/pdf/1605.08814}
}

@article{valivarthi2020Teleportation,
  title = {Teleportation Systems Toward a Quantum Internet},
  author = {Valivarthi, Raju and Davis, Samantha I. and Pe\~na, Cristi\'an and Xie, Si and Lauk, Nikolai and Narv\'aez, Lautaro and Allmaras, Jason P. and Beyer, Andrew D. and Gim, Yewon and Hussein, Meraj and Iskander, George and Kim, Hyunseong Linus and Korzh, Boris and Mueller, Andrew and Rominsky, Mandy and Shaw, Matthew and Tang, Dawn and Wollman, Emma E. and Simon, Christoph and Spentzouris, Panagiotis and Oblak, Daniel and Sinclair, Neil and Spiropulu, Maria},
  journal = {PRX Quantum},
  volume = {1},
  pages = {020317},
  numpages = {16},
  year = {2020},
  month = {Dec},
  publisher = {American Physical Society},
  doi = {10.1103/PRXQuantum.1.020317},
  url = {https://link.aps.org/doi/10.1103/PRXQuantum.1.020317}
}

@article{shen2023hertz,
  title={Hertz-rate metropolitan quantum teleportation},
  author={Si Shen and Chenzhi Yuan and Zichang Zhang and Hao Yu and Ruiming Zhang and Chuanrong Yang and Hao Li and Zhen Wang and You Wang and Guangwei Deng and Haizhi Song and Lixing You and Yunru Fan and Guangcan Guo and Qiang Zhou},
  volume={12},
  pages={115},
  journal={Light: Science \& Applications},
  year={2023},
  url={https://www.nature.com/articles/s41377-023-01158-7}
}

@article{gisin2003longdistance,
  title={Long-distance teleportation of qubits at telecommunication wavelengths},
  author={Marcikic, I and de Riedmatten, Hugues and Tittel, Wolfgang and Zbinden, Hugo and Gisin, Nicolas},
  journal={Nature},
  volume={421},
  pages={509--513},
  year={2003},
  doi={10.1038/nature01376},
  url={https://www.nature.com/articles/nature01376}
}

@article{bussiere2014telecom,
  title={Quantum teleportation from a telecom-wavelength photon to a solid-state quantum memory},
  author={Bussières, Félix and Clausen, Christoph and Tiranov, Alexey and Korzh, Boris and Verma, Varun B. and Nam, Sae Woo and Marsili, Francesco and Ferrier, Alban and Goldner, Philippe and Herrmann, Harald and Silberhorn, Christine and Sohler, Wolfgang and Afzelius, Mikael and Gisin, Nicolas},
  journal={Nature Photonics},
  volume={8},
  number={10},
  pages={775--778},
  year={2014},
  doi={10.1038/nphoton.2014.215},
  url={https://www.nature.com/articles/nphoton.2014.215}
}

@article{kucera2024demonstration,
  title={Demonstration of quantum network protocols over a 14‑km urban fiber link},
  author={Stephan Kucera and Christian Haen and Elena Arenskoetter and Tobias Bauer and Jonas Meiers and Marlon Schaefer and Ross Boland and Milad Yahyapour and Maurice Lessing and Ronald Holzwarth and Christoph Becher and Juergen Eschner},
  journal={npj Quantum Information},
  volume={10},
  pages={88},
  year={2024},
  doi={10.1038/s41534-024-00886-x},
  url={https://www.nature.com/articles/s41534-024-00886-x}
}

@article{main2025distributed,
  title={Distributed quantum computing across an optical network link},
  author={D. Main and P. Drmota and D. P. Nadlinger and E. M. Ainley and A. Agrawal and B. C. Nichol and R. Srinivas and G. Araneda and D. M. Lucas},
  journal={Nature},
  volume={638},
  pages={383--388},
  year={2025},
  publisher={Nature Publishing Group UK London},
  doi = {10.1038/s41586-024-08404-x}
}

@article{jahromi2022remote,
  title={Remote sensing and faithful quantum teleportation through non-localized qubits},
  author={Jahromi, Hossein Rangani},
  journal={Physics Letters A},
  volume={424},
  pages={127850},
  year={2022},
  publisher={Elsevier},
  doi = {10.1016/j.physleta.2021.127850}
}

@article{kimble2008quantum,
  title={The quantum internet},
  author={Kimble, H Jeff},
  journal={Nature},
  volume={453},
  number={7198},
  pages={1023--1030},
  year={2008},
  publisher={Nature Publishing Group},
  doi = {10.1038/nature07127}
}

@article{briegel98prl,
  title = {Quantum Repeaters: The Role of Imperfect Local Operations in Quantum Communication},
  author = {H.-J. Briegel and W. D\"{u}r and J. I. Cirac and P. Zoller},
  journal = {Phys. Rev. Lett.},
  volume = {81},
  issue = {26},
  pages = {5932--5935},
  year = {1998},
  month = {Dec},
  publisher = {American Physical Society},
  doi = {10.1103/PhysRevLett.81.5932},
  url = {https://link.aps.org/doi/10.1103/PhysRevLett.81.5932}
}

@article{sena25JOCN,
author = {Matheus Sena and Mael Flament and Shane Andrewski and Ioannis Caltzidis and Niccol\`{o} Bigagli and Thomas Rieser and Gabriel Bello Portmann and Rourke Sekelsky and Ralf-Peter Braun and Alexander N. Craddock and Maximilian Schulz and Klaus D. J\"{o}ns and Michaela Ritter and Marc Geitz and Oliver Holschke and Mehdi Namazi},
journal = {J. Opt. Commun. Netw.},
number = {12},
pages = {1072--1081},
publisher = {Optica Publishing Group},
title = {High-fidelity quantum entanglement distribution in metropolitan fiber networks with co-propagating classical traffic},
volume = {17},
month = {Dec},
year = {2025},
url = {https://opg.optica.org/jocn/abstract.cfm?URI=jocn-17-12-1072},
doi = {10.1364/JOCN.575396},
}

@inproceedings{sena25OFC,
author = {Matheus Sena and Mael Flament and Mehdi Namazi and Shane Andrewski and Gabriel Portmann and Ralf-Peter Braun and Marwa Youssef-Sayed and Ronny D\"{o}ring and Michaela Ritter and Oliver Holschke and Marc Geitz},
booktitle = {Optical Fiber Communication Conference (OFC) 2025},
journal = {Optical Fiber Communication Conference (OFC) 2025},
pages = {M4E.2},
publisher = {Optica Publishing Group},
title = {High-Fidelity Entanglement Distribution Through Berlin Using an Operator's Fiber Infrastructure},
year = {2025},
url = {https://opg.optica.org/abstract.cfm?URI=OFC-2025-M4E.2},
doi = {10.1364/OFC.2025.M4E.2},
}

@article{Craddock24PRX,
  title = {Automated Distribution of Polarization-Entangled Photons Using Deployed New York City Fibers},
  author = {Craddock, Alexander N. and Lazenby, Anne and Portmann, Gabriel Bello and Sekelsky, Rourke and Flament, Mael and Namazi, Mehdi},
  journal = {PRX Quantum},
  volume = {5},
  issue = {3},
  pages = {030330},
  numpages = {7},
  year = {2024},
  month = {Aug},
  publisher = {American Physical Society},
  doi = {10.1103/PRXQuantum.5.030330},
  url = {https://link.aps.org/doi/10.1103/PRXQuantum.5.030330}
}

@misc{Flament2025AU2021365730B2,
  author       = {M. Flament and M. Namazi and G. B. Portmann and R. Sekelsky},
  title        = {Systems and methods for real-time polarization drift compensation in optical fiber channels used for quantum communications},
  howpublished = {Patent US12361308B2},
  year         = {2025},
  note         = {Available: \url{https://patents.google.com/patent/US12361308B2/}}
}

@article{Bennett95,
  title = {Teleporting an unknown quantum state via dual classical and Einstein-Podolsky-Rosen channels},
  author = {C. H. Bennett and G. Brassard and C. Cr\'{e}peau and R. Jozsa and A. Peres and W. K. Wootters},
  journal = {Phys. Rev. Lett.},
  volume = {70},
  issue = {13},
  pages = {1895--1899},
  year = {1993},
  month = {Mar},
  publisher = {American Physical Society},
  doi = {10.1103/PhysRevLett.70.1895},
  url = {https://link.aps.org/doi/10.1103/PhysRevLett.70.1895}
}

@inproceedings{Rieser2025ECOC_OBandEntanglement,
  title     = {Distributing, Routing and Multiplexing O-Band Polarization-Entangled Photons with C-Band Classical Light over an Operator’s Metropolitan Fiber Network},
  author    = {Rieser, Thomas and Sena, Matheus and Andrewski, Shane and Bigagli, Niccol{\`o} and Flament, Mael and Braun, Ralf-Peter and Ritter, Michaela and Namazi, Mehdi and Geitz, Marc},
  booktitle = {Proceedings of the European Conference on Optical Communication (ECOC)},
  year      = {2025},
  address   = {Copenhagen, Denmark}
}

@article{legero2003time,
  title={Time-resolved two-photon quantum interference},
  author={Legero, Thomas and Wilk, Tatjana and Kuhn, Axel and Rempe, Gerhard},
  journal={Applied Physics B},
  volume={77},
  number={8},
  pages={797--802},
  year={2003},
  publisher={Springer}
}

@article{FanPRA23,
  title = {Energy-time entanglement coexisting with fiber-optical communication in the telecom $C$ band},
  author = {Y.-R. Fan and Y. Luo and Z.-C. Zhang and Y.-B. Li and S. Liu and D. Wang and D.-C. Zhang and G.-W. Deng and Y. Wang and H.-Z. Song and Z. Wang and L.-X. You and C.-Z. Yuan and G.-C. Guo and Q. Zhou},
  journal = {Phys. Rev. A},
  volume = {108},
  issue = {2},
  pages = {L020601},
  numpages = {5},
  year = {2023},
  month = {Aug},
  publisher = {American Physical Society},
  doi = {10.1103/PhysRevA.108.L020601},
  url = {https://link.aps.org/doi/10.1103/PhysRevA.108.L020601}
}

@article{james2001measurement,
  title = {Measurement of qubits},
  author = {James, Daniel F. V. and Kwiat, Paul G. and Munro, William J. and White, Andrew G.},
  journal = {Phys. Rev. A},
  volume = {64},
  issue = {5},
  pages = {052312},
  numpages = {15},
  year = {2001},
  month = {Oct},
  publisher = {American Physical Society},
  doi = {10.1103/PhysRevA.64.052312},
  url = {https://link.aps.org/doi/10.1103/PhysRevA.64.052312}
}

@online{swaperooboubou,
  author       = {Alexander N. Craddock and Tyler Cowan and Niccol\`{o} Bigagli and Suresh Yekasiri and Dylan Robinson and Gabriel Bello Portmann and Ziyu Guo and Michael Kilzer and Jiapeng Zhao and Mael Flament and Javad Shabani and Reza Nejabati and Mehdi Namazi},
  title        = {High-rate Scalable Entanglement Swapping Between Remote Entanglement Sources on Deployed New York City Fibers},
  year         = {2026},
  eprint       = {arXiv:2602.15653},
  eprinttype   = {arXiv},
  eprintclass  = {quant-ph},
  url          = {https://arxiv.org/abs/2602.15653},
  note         = {Submitted 17 Feb 2026}
}

@article{lagorivera2023longdistance,
  title={Long distance multiplexed quantum teleportation from a telecom photon to a solid-state qubit},
  author={Lago-Rivera, Dario and Rakonjac, Jelena V. and Grandi, Samuele and de Riedmatten, Hugues},
  journal={Nature Communications},
  volume={14},
  pages={1889},
  year={2023},
  month={April},
  doi={10.1038/s41467-023-37518-5},
  publisher={Nature Publishing Group},
  url={https://www.nature.com/articles/s41467-023-37518-5}
}

@article{haenni25heralded,
  title     = {Heralded Entanglement of On-Demand Spin-Wave Solid-State Quantum Memories for Multiplexed Quantum Network Links},
  author    = {H\"anni, Jonathan and Rodr\'iguez-Moldes, Alberto E. and Appas, F\'elicien and Wengerowsky, Soeren and Lago-Rivera, Dario and Teller, Markus and Grandi, Samuele and de Riedmatten, Hugues},
  journal   = {Phys. Rev. X},
  volume    = {15},
  issue     = {4},
  pages     = {041003},
  numpages  = {},
  year      = {2025},
  month     = {10},
  publisher = {American Physical Society},
  doi       = {10.1103/wvv1-6lg8},
  url       = {https://doi.org/10.1103/wvv1-6lg8}
}

@article{song2026hollowcore,
  title   = {Quantum teleportation over a field-deployed hollow-core fibre network},
  author  = {Song, Ri-Yao and Zhao, Ya-Zhou and Fan, Yun-Ru and Ma, Yang-Bin and Wei, Yan-Yu and Shen, Si and Li, Hao and You, Li-Xing and Guo, Kai and Guo, Guang-Can and Zhou, Qiang},
  journal = {arXiv preprint arXiv:2607.25352},
  year    = {2026},
  eprint  = {2607.25352},
  archivePrefix = {arXiv},
  primaryClass  = {quant-ph},
  doi     = {10.48550/arXiv.2607.25352}
}

@article{craddock2019,
  title = {Quantum Interference between Photons from an Atomic Ensemble and a Remote Atomic Ion},
  author = {Craddock, A. N. and Hannegan, J. and Ornelas-Huerta, D. P. and Siverns, J. D. and Hachtel, A. J. and Goldschmidt, E. A. and Porto, J. V. and Quraishi, Q. and Rolston, S. L.},
  journal = {Phys. Rev. Lett.},
  volume = {123},
  issue = {21},
  pages = {213601},
  numpages = {6},
  year = {2019},
  month = {Nov},
  publisher = {American Physical Society},
  doi = {10.1103/PhysRevLett.123.213601},
  url = {https://link.aps.org/doi/10.1103/PhysRevLett.123.213601}
}
\newpage

% \section{Appendix}

% Reconstructed density matrices for the teleported output states in the 30-km configuration without co-propagating classical traffic, displayed as real (left) and imaginary (right) components in the ${\ket{H},\ket{V}}$ basis. The measured output states ${\ket{V},\ket{A},\ket{R}}$ correspond to the input states ${\ket{H},\ket{D},\ket{R}}$, as expected in the absence of the unitary correction operation in the case of $\ket{\Psi^-}$ conditioning.

% \begin{figure}[htbp]
%  \centering
%  \includegraphics[width=1\columnwidth]{figures/density_matrices_local.pdf}
%  \caption{Reconstructed density
% matrices $\rho$ (real and imaginary parts in the $\{\ket{H},\ket{V}\}$ basis from single-state tomography for the local configuration.}
%  \label{fig:tomography}
% \end{figure}

% \begin{figure}[htbp]
%  \centering
%  \includegraphics[width=1\columnwidth]{figures/density_matrices_30km.pdf}
%  \caption{Reconstructed density
% matrices $\rho$ (real and imaginary parts in the $\{\ket{H},\ket{V}\}$ basis from single-state tomography for the 30 km configuration without copropagating classical traffic.}
%  \label{fig:tomography}
% \end{figure}

% \begin{figure}[htbp]
%  \centering
%  \includegraphics[width=1\columnwidth]{figures/density_matrices_coprop.pdf}
%  \caption{Reconstructed density
% matrices $\rho$ (real and imaginary parts in the $\{\ket{H},\ket{V}\}$ basis from single-state tomography for the 30 km configuration with copropagating classical traffic.}
%  \label{fig:tomography}
% \end{figure}

\end{document}